\documentclass[12pt]{article}
\usepackage{epsfig,a4}
\def\be{\begin{equation}}
\def\ee{\end{equation}}
\def\bea{\begin{eqnarray}}
\def\eea{\end{eqnarray}}
\def\rarr{\rightarrow}
\def\del{\partial}
\def\kf{{\bf k}}
\def\qf{{\bf q}}
\def\lf{{\bf l}}
\def\N{{\rm I\kern-.18em N}} 
\def\nn{\nonumber}
\def\fr{\frac}

\newdimen\picraise
\newcommand\picbox[1]
{
  \setbox0=\hbox{\input{#1}}
  \picraise=-0.5\ht0 
  \advance\picraise by 0.5\dp0
  \advance\picraise by 3pt     
  \hbox{\raise\picraise \box0}
}
\begin{document}
\begin{titlepage}
\begin{flushright}
HD-THEP-01-16\\
hep-ph/0105181
\end{flushright}
\vfill
\vspace*{1cm}
\begin{center}
\boldmath
{\Large{\bf Conformal Invariance of Unitarity Corrections $^*$}}
\unboldmath
\end{center}
\vspace{1.2cm}
\begin{center}
{\bf \large 
Carlo Ewerz
}
\end{center}
\begin{center}
{\sl
Institut f\"ur Theoretische Physik, Universit\"at Heidelberg\\
Philosophenweg 16, D-69120 Heidelberg, Germany}
\end{center}
\vfill
\begin{abstract}
We study perturbative unitarity corrections in the 
generalized leading logarithmic approximation in high 
energy QCD. It is shown that the corresponding amplitudes 
with up to six gluons in the $t$-channel are conformally 
invariant in impact parameter space. In particular we 
give a new representation for the two--to--six reggeized 
gluon vertex in terms of conformally invariant functions.
With the help of this representation an interesting regularity 
in the structure of the two--to--four and the two--to--six 
transition vertices is found. 
\end{abstract}
\vfill
\vspace{5em}
\hrule width 5.cm
\vspace*{.5em}
{\small \noindent 
$^*$Work supported in part by the EU Fourth Framework Programme
`Training and Mobility of Researchers', Network `Quantum Chromodynamics
and the Deep Structure of Elementary Particles',
contract FMRX-CT98-0194 (DG 12 - MIHT).
}
\end{titlepage}

\section{Introduction}
\label{intro}

One of the most interesting properties of the leading logarithmic 
approximation (LLA) in the Regge limit of QCD is 
the conformal invariance of scattering amplitudes in two--dimensional 
impact parameter space. In LLA  
the scattering of small color dipoles is described by the BFKL Pomeron 
\cite{FKL,BL}. It resums large logarithms of the energy $\sqrt{s}$ 
which can compensate the smallness of the strong coupling constant 
$\alpha_s$. After a Fourier transformation from transverse momentum 
space to two--dimensional impact parameter space the corresponding 
amplitude is invariant under global conformal transformations, i.\,e.\ 
under M\"obius transformations \cite{Lip86}. 
The perturbative (BFKL) Pomeron describes the exchange of 
two interacting reggeized gluons in the $t$-channel. 
The unitarization of the scattering amplitude requires to take 
into account also contributions with larger numbers of gluons 
in the $t$-channel which are consequently called 
unitarity corrections. 
In \cite{Lipngluon} it was shown that 
also a system of $n$ interacting reggeized gluons (described 
by the BKP equations \cite{Bartelskernels,BKP}) 
is invariant under conformal transformations in impact 
parameter space. 

More recently also transitions 
between states containing different numbers of reggeized 
gluons were studied 
\cite{BPLB}--\cite{CE}. 
These transition vertices have been obtained in the 
so--called generalized leading logarithmic approximation 
(GLLA) \cite{Bartelskernels,Bartelsnuclphys,Bartelsinteq}. 
Amplitudes with up to six reggeized gluons in the $t$-channel 
have been investigated in the GLLA. It was found that these 
amplitudes have the structure of an effective field theory in which only 
states with even numbers of gluons occur. These states are coupled to 
each other via number changing vertices. In particular the 
two--to--four gluon vertex $V_{2\rarr 4}$ \cite{BZPhys,BW} 
and the two--to--six gluon vertex $V_{2\rarr 6}$ \cite{BE} have 
been calculated explicitly. These two are the only number changing 
vertices present in the amplitudes with up to six gluons. 

The natural question arises whether also the transition vertices 
exhibit conformal symmetry in impact parameter space. For the 
case of the two--to--four gluon vertex a positive answer to this question 
was given in \cite{BLW}. The proof of conformal invariance of 
this vertex was later simplified in \cite{Braun,Vacca}. 
In the present paper we will show that also the two--to--six 
gluon vertex is conformally invariant. 
We can then further conclude that all amplitudes describing 
perturbative unitarity corrections with up to six gluons 
in the $t$-channel are conformally invariant. 
This becomes clear when we recall the field theory structure 
observed in those amplitudes. The amplitudes in fact contain 
only elements which are conformally invariant. The amplitude 
describing the production of three gluons, for example, turns out 
to be a superposition of two--gluon (BFKL) amplitudes \cite{BW}. 
Similarly, the five--gluon amplitudes is a superposition of 
two-- and four--gluon amplitudes \cite{BE} etc. 
The conformal invariance of the building blocks, i.\,e.\ of the 
$n$-gluon states and of the transition vertices, is thus sufficient 
to prove the conformal invariance of the full amplitudes. 

In section \ref{vg} we briefly review the BFKL kernel 
and the two--to--four reggeized gluon vertex 
as well as an auxiliary function which is already known to be 
conformally invariant. In section \ref{six} we then show 
that the two--to--six reggeized gluon vertex can be 
represented in terms of this function. The comparison 
of the new representation for  $V_{2\rarr 6}$ with a 
similar one for $V_{2\rarr 4}$ allows us to observe 
an interesting regularity in the structure of the number changing 
vertices. 

\section{The BFKL kernel and the two-to-four gluon vertex}
\label{vg}

The interaction of two reggeized gluons is described by the 
BFKL kernel. In impact parameter space it can be represented 
most conveniently using pseudodifferential operators. One 
defines complex notation for the two--dimensional gluon 
coordinates, 
\be
\label{complcoord}
  \vec{\rho} = (\rho_x,\rho_y) \;\;\; \longrightarrow \;\;\;
\rho= \rho_x+i \rho_y
\ee
and corresponding derivatives $\del=\frac{\del}{\del \rho}$.
The interaction kernel for two gluons with complex coordinates 
$\rho_1$ and $\rho_2$ then reads \cite{LipMueller,Liprep} 
\be
\label{holomorph}
 {\cal K} = \frac{g^2 N_c}{8 \pi^2} (K +K^*)
\ee
with 
\be
\label{kernelk}
K = \log[(\rho_1-\rho_2)^2 \del_1] + \log[(\rho_1-\rho_2)^2 \del_2] 
   - 2 \log (\rho_1-\rho_2) - 2 \psi(1) 
\,.
\ee
Here $\psi$ denotes the logarithmic derivative of the Euler gamma 
function. 

According to (\ref{holomorph}) the BFKL kernel is a sum of 
two operators, the first of which acts only on the holomorphic 
coordinates $\rho_1,\rho_2$, whereas the second acts on 
the antiholomorphic coordinates $\rho_1^*,\rho_2^*$ only. 
This property of the BFKL kernel is called holomorphic separability. 

With the help of the representation (\ref{holomorph}) it is 
relatively easy to show that the kernel ${\cal K}$ is invariant 
under conformal (M\"obius) transformations of the gluon 
coordinates 
\be
\label{Moebius}
 \rho \longrightarrow \frac{a \rho + b}{c\rho +d} \,, 
\mbox{\hspace{2cm}}
\rho^* \longrightarrow \frac{a^* \rho^* + b^*}{c^*\rho^* +d^*} 
\ee
with $a d- b c=a^* d^* -b^* c^* =1$. These transformations are thus 
characterized by 
\be
  \left(
  \begin{array}{cc}
  {a}&{b}\\
  {c}&{d}
  \end{array}
  \right)
  \in SL(2,{\bf{C}}) / Z_2 \,,
\ee
i.\,e.\ the group of projective conformal transformations. 

Before we consider the two--to--four gluon 
transition vertex $V_{2 \rightarrow 4}$ we define 
some useful functions. Let $\phi_2(\kf_1,\kf_2)$ denote 
a two--gluon (BFKL) amplitude, 
$\kf_i$ being the transverse momenta of the gluons. 
Then we define 
\bea
a(\kf_1,\kf_2,\kf_3) &=&  \int \fr{d^2\lf}{(2 \pi)^3}  
       \fr{\kf_1^2}{(\lf-\kf_2)^2 [\lf-(\kf_1+\kf_2)]^2}
      \,\phi_2\!\left(\lf, \sum_{j=1}^{3}\kf_j-\lf \right) \,,
\\
b(\kf_1,\kf_2) &=& a(\kf_1,\kf_2,\kf_3=0)\,, 
\\
c(\kf_1) &=& b(\kf_1,\kf_2=0)\,,
\\
s(\kf_1,\kf_2,\kf_3) &=& 
- \fr{2}{N_c g^2} \beta(\kf_1)  \phi_2(\kf_1+\kf_2,\kf_3) \,,
\\
t(\kf_1,\kf_2) &=& s(\kf_1,\kf_3=0,\kf_2) \,,
\eea
where $\alpha=1+\beta$ is the well--known gluon trajectory function with 
\be
\label{traject}
  \beta(\kf^2) = - \frac{N_c}{2} g^2  \int \frac{d^2\lf}{(2 \pi)^3} 
          \frac{\kf^2}{\lf^2 (\lf -\kf)^2} 
\,.
\ee
These functions are not infrared finite by themselves but will occur 
only in infrared finite combinations. An important example for 
such an infrared finite combination is the function $G$, 
\bea
\label{gdef}
 G(\kf_1,\kf_2,\kf_3) &=& \frac{g^2}{2}\, [
  \,2 c(123) - 2 b(12,3) - 2 b(23,1) + 2 a(2,1,3) \nn \\ 
 & &  \hspace{.6cm} +\, t(12,3) + t(23,1) - s(2,1,3) - s(2,3,1) ] 
\,,
\eea
$g$ being the gauge coupling. 
Here we have introduced a shorthand notation for the 
momentum arguments by replacing the momentum $\kf_i$ by 
its index $i$. A string of indices stands for a sum of the 
corresponding momenta, for example we have 
$t(12,3)=t(\kf_1+\kf_2,\kf_3)$. 
The function $G$ describes a transition kernel from the two gluons in the 
amplitude $\phi_2$ to three gluons with momenta $\kf_1,\kf_2,\kf_3$. 
The functions $a,b,c$ correspond to real gluon emission, 
whereas $s$ and $t$ describe virtual corrections. 
Using the definitions of these functions it
 is also possible to write $G=\widetilde{G} \phi_2$ with an integral 
operator $\widetilde{G}$ acting on the two--gluon amplitude $\phi_2$. 
We note that the function $G(\kf_1,\kf_2,\kf_3)$ 
is not symmetric in its three arguments. 
It vanishes when the first or the last argument vanishes, 
\be
\label{gwird0}
G(\kf_1=0 ,\kf_2,\kf_3) = G(\kf_1,\kf_2,\kf_3=0) = 0 \,,
\ee
but it does not vanish when its second argument $\kf_2$ vanishes. 
Interestingly, in this case $G$ happens to reduce (up to a 
color factor) to a BFKL 
kernel ${\cal K}$, 
\bea
\label{gwirdbfkl}
G(\kf_1,\kf_2=0,\kf_3) &=& 
\frac{1}{N_c}{\cal K}(\kf_1,\kf_3) \otimes \phi_2 \\
&=& \frac{g^2}{2} [ 2 c(13) - 2 b(1,3) - 2 b(3,1) +t(1,3) + t(3,1)]
\,.
\eea
In the unitarity corrections $G$ never occurs as an isolated 
object. It should therefore not be confused with one of the transition 
vertices of the effective field theory of unitarity corrections. 
The transition vertices, however, can be expressed in terms 
of $\widetilde{G}$, as we will show explicitly for 
$V_{2 \rightarrow 4}$ and $V_{2 \rightarrow 6}$. 

Let us now turn to the two--to--four gluon transition vertex 
$V_{2 \rightarrow 4}$. It couples a two--gluon amplitude $\phi_2$ 
to a four--gluon amplitude $\phi_4$. We will assume that 
the two gluons in $\phi_2$ are in a color singlet state. 
(The vertex $V_{2 \rightarrow 4}$ is in fact known 
only for this situation.) 
The arguments of $V_{2 \rightarrow 4}$ are the 
transverse momenta $\qf_j$ 
of the two incoming gluons and the transverse 
momenta $\kf_i$ of the four outgoing gluons. 
In addition, 
the transition vertex carries four gluon color labels for 
the outgoing gluons. The vertex was computed in \cite{BZPhys,BW} 
where it was shown to have the following structure, 
\bea
\label{colV}
 V_{2 \rightarrow 4}^{a_1a_2a_3a_4}(\{\qf_j\};\kf_1,\kf_2,\kf_3,\kf_4) 
 \!&=&\! 
  \delta_{a_1a_2} \delta_{a_3a_4} 
V(\{\qf_j\};\kf_1,\kf_2;\kf_3,\kf_4)
\nonumber \\
& & 
 +\, \delta_{a_1a_3} \delta_{a_2a_4} 
V(\{\qf_j\};\kf_1,\kf_3;\kf_2,\kf_4)
 \nonumber \\
 & &
 +\,\delta_{a_1a_4} \delta_{a_2a_3} 
V(\{\qf_j\};\kf_1,\kf_4;\kf_2,\kf_3)
\,.
\eea
As can be derived from this color and momentum structure, 
the vertex is completely symmetric in the four outgoing gluons, i.\,e.\ 
under the simultaneous exchange of color labels and momenta 
of the gluons. 
The vertex $V_{2 \rightarrow 4}$ should actually be understood 
as an integral operator acting on $\phi_2$. We will be mainly 
interested in the combination $V_{2 \rightarrow 4}\phi_2$, 
which is its action on the two--gluon amplitude. 
The function $V$ and its action on $\phi_2$ were 
first computed in terms of integrals of the form $a,b,c,s$, and $t$. 
We will not need this representation explicitly here 
and give an alternative expression momentarily. 
We note that the function 
$V(\{\qf_j\};\kf_1,\kf_2;\kf_3,\kf_4)$
is symmetric in the two momenta $\qf_j$, in the two 
momenta $\kf_1$ and $\kf_2$, as well as in 
$\kf_3$ and $\kf_4$, hence their separation in the notation. 
It is further symmetric under the exchange of the pairs 
$\{ \kf_1,\kf_2\}$ and $\{ \kf_3,\kf_4\}$. 

To explain the conformal invariance of $V_{2 \rightarrow 4}$ 
we consider an amplitude ${\cal A}_4$ which is the convolution 
of the vertex $V_{2 \rightarrow 4}$ with 
the BFKL amplitude $\phi_2$ and a four--gluon state $\phi_4$. 
The latter is assumed to be a solution of 
the four--particle BKP equation. The amplitude ${\cal A}_4$ 
has the form 
\bea
\label{amplia4}
{\cal A}_4 &=& 
\int \prod_{j=1}^2 d^2\qf_j \prod_{i=1}^4 d^2\kf_i
\,\phi_2(\qf_1,\qf_2)  \,
V_{2 \rightarrow 4}^{a_1a_2a_3a_4}
(\{ {\bf q}_j\},\kf_1,\kf_2,\kf_3,\kf_4) 
\times \nn \\
&& \quad \times \,
\phi_4^{a_1a_2a_3a_4}(\kf_1,\kf_2,\kf_3,\kf_4) 
\,\delta\left( \,\sum_{j=1}^2 \qf_j - \sum_{i=1}^4 \kf_i \right)
\,.
\eea
Transverse momentum space and impact parameter space 
are related to each other by a Fourier transformation, 
\bea
\label{fourier2}
\phi_2(\qf_1,\qf_2)  &=& 
\int \prod_{j=1}^2  
\left[ d^2\rho_{j'} \,e^{i \qf_{j}\rho_{j'}} \right]
\phi_2(\rho_{1'},\rho_{2'}) \\
\phi_4(\kf_1,\kf_2,\kf_3,\kf_4)  &=& 
\int \prod_{i=1}^4  
\left[ d^2\rho_i \,e^{-i \kf_i\rho_i} \right]
\phi_4(\rho_1,\rho_2,\rho_3,\rho_4) 
\label{fourier4}
\,,
\eea
where the gluon coordinates $\rho$ are understood in complex notation 
as introduced in eq.\ (\ref{complcoord}). 
Applying this procedure to (\ref{amplia4}) 
defines the Fourier transform of the transition 
vertex. Using the conformal invariance of the functions 
$\phi_2$ and $\phi_4$ (see \cite{Lipngluon}) 
one can prove the invariance of 
the whole amplitude ${\cal A}_4$ under a simultaneous 
M\"obius transformation of all gluon coordinates $\rho_i$ and $\rho_{j'}$ 
according to (\ref{Moebius}), and then infer that the vertex 
$V_{2 \rightarrow 4}$ is in fact conformally invariant. 
In \cite{BLW} this proof was performed using the original 
representation for $V\phi_2$ mentioned above. 
In \cite{Braun,Vacca} it was shown that the proof of 
conformal invariance can be simplified by making use of a 
representation that can be found already in \cite{BW} for the 
forward direction ($\sum_i \kf_i = 0$), 
\bea
\label{vmitg}
  (V \phi_2 )(\kf_1,\kf_2;\kf_3,\kf_4) &=& \frac{g^2}{2} [ \, 
       G(12,-,34) \\
 & & \hspace{.5cm} 
  -\, G(12,3,4) - G(12,4,3) - G(1,2,34) - G(2,1,34) \nn \\
 & & \hspace{.5cm} 
  + \,G(1,23,4) + G(2,13,4) + G(1,24,3) + G(2,14,3)] \nn 
\,.
\eea
This representation holds for the non--forward direction as well. 
After Fourier transformation to configuration space, 
the function $G$ was shown to be conformally invariant by itself. 
The proof uses essentially the same method as the original one 
for $V$ in \cite{BLW}, now applied to the simpler function $G$. 
Again one defines an amplitude ${\cal A}_3$ in analogy to 
${\cal A}_4$ in (\ref{amplia4}), now with a three--gluon 
amplitude $\phi_3$ instead of $\phi_4$ and without the 
contraction in color space. The Fourier transform is then 
defined in analogy to (\ref{fourier2}) and (\ref{fourier4}). 
The most difficult step is to show the invariance under 
inversion, $\rho \to 1/\rho$. The explicit proof is rather 
technical and requires a careful treatment of the necessary 
regularization parameters (for details see \cite{Braun,Vacca,BLW}). 
A subtle point is the possible occurrence of infrared logarithms 
which can potentially break the conformal invariance of transition 
kernels in impact parameter space. Such logarithms can 
arise if the kernels do not vanish when their momentum arguments 
tend to zero. 
The function $G$ does indeed not vanish when 
its second momentum argument vanishes, 
as we have seen in eq.\ (\ref{gwirdbfkl}), and one 
could thus expect dangerous infrared logarithms to arise. 
But it turns out that in the unitarity corrections $G$ occurs only in 
combinations which have the property to vanish for 
vanishing momentum arguments. The function 
$V \phi_2$ in the two--to--four gluon vertex 
for example is such a combination, see eq.\ (\ref{vmitg}). 
One can easily derive that 
\be
\label{nullstellv}
\left.
(V \phi_2) (\kf_1,\kf_2;\kf_3,\kf_4) 
\right|_{\kf_i = 0} = 0  \;\;\;\;\;\;  (i\in\{1,\dots,4\}) 
\,.  
\ee
This implies that the potentially dangerous logarithms 
cancel in the vertices of the effective field theory of 
unitarity corrections and do not need to be considered 
any further. With this additional piece of 
information the conformal invariance of the vertex 
$V_{2 \rightarrow 4}$ can then be derived easily because 
it is a superposition of conformally invariant functions 
$G$. 

The explicit expression for the function $G$ in impact parameter 
space can be found in \cite{Braun,Vacca} and will not be needed here. 
Unfortunately, it was so far not possible to find a representation 
for $G$ which is of similar simplicity as (\ref{kernelk}) for the 
BFKL kernel, namely in terms of pseudodifferential operators. 
In particular, it appears that $G$ and the 
vertex $V_{2 \rightarrow 4}$ actually lack the property of 
holomorphic separability \cite{Hans}. 

\boldmath
\section{The two-to-six gluon vertex}
\unboldmath
\label{six}

We now turn to the two--to--six gluon vertex derived in \cite{BE}. 
Again we consider its action on a two--gluon state $\phi_2$. 
Let $\kf_i$ be the transverse momenta of the six produced gluons, 
and let $a_i$ denote the corresponding color labels. 
Then the vertex has the form 
\be
 (V_{2 \rightarrow 6}^{a_1a_2a_3a_4a_5a_6}\phi_2)
(\kf_1,\kf_2,\kf_3,\kf_4,\kf_5,\kf_6) = 
\sum   d_{a_1a_2a_3} d_{a_4a_5a_6} (W \phi_2)(1,2,3;4,5,6) 
\,,
\label{newpiece}
\ee
where $d_{abc}$ denotes the symmetric structure constant 
of $su(N_c)$, and we again use the shorthand notation for 
the momentum arguments introduced below eq.\ (\ref{gdef}). 
The sum extends over all (ten) partitions of the six gluons 
into two groups containing three gluons each, 
\bea
\label{explsumoverw}
\lefteqn{\sum   d_{a_1a_2a_3} d_{a_4a_5a_6} (W \phi_2)(1,2,3;4,5,6) =}\nn\\
&=& d_{a_1a_2a_3} d_{a_4a_5a_6} (W \phi_2)(1,2,3;4,5,6) 
\nn \\
&&+ \, d_{a_1a_2a_4} d_{a_3a_5a_6} (W \phi_2)(1,2,4;3,5,6) 
+ \dots 
\nn \\
&&
+ \,d_{a_1a_5a_6} d_{a_2a_3a_4} (W \phi_2)(1,5,6;2,3,4) 
\,.
\eea
The symmetry properties of the vertex are very similar to those 
of the two--to--four vertex $V_{2 \rightarrow 4}$ discussed 
earlier. 
The function $(W \phi_2)(1,2,3;4,5,6)$ is completely symmetric in the first 
three gluon momenta as well as in the last three momenta, hence 
the notation. 
Further, $W \phi_2$ is symmetric under the exchange 
of its first three with its last three arguments, 
\be
(W \phi_2)(\kf_1,\kf_2, \kf_3;\kf_4,\kf_5,\kf_6) = 
(W \phi_2)(\kf_4,\kf_5, \kf_6;\kf_1,\kf_2,\kf_3)
\,.
\ee
From (\ref{newpiece}) then follows the symmetry of the full 
vertex $V_{2 \rightarrow 6}$ 
in the six gluons, i.\,e.\ the symmetry under the simultaneous 
exchange of momenta and color labels of the gluons. 
In \cite{BE} the function $W\phi_2$ was calculated in terms 
of the integrals $a,b,c,s$, and $t$. The corresponding expression 
is rather long and we do not give it explicitly here. 

The transformation to impact parameter space is performed 
as in the case of $V_{2 \rightarrow 4}$. An amplitude 
${\cal A}_6$ is defined similarly to ${\cal A}_4$ in 
eq.\ (\ref{amplia4}). Now we have $V_{2 \rightarrow 6}$ 
instead of $V_{2 \rightarrow 4}$, and $\phi_4$ in that 
equation is replaced by a conformally invariant state $\phi_6$ 
of six reggeized gluons solving the six-reggeon BKP equation. Obviously 
the integration is extended to all six momenta $\kf_i$. 
The Fourier transformation is then performed in analogy to 
eqs.\ (\ref{fourier2}) and (\ref{fourier4}) to get the vertex in impact 
parameter space. The most convenient way to prove its 
invariance under conformal transformations (\ref{Moebius}) 
is again to show that it can be written as a sum of $G$-functions 
already in momentum space, and thus also in impact 
parameter space. Such a representation for 
$W \phi_2$ can in fact be found,   
\bea
\label{wmitg}
\lefteqn{(W\phi_2)(\kf_1,\kf_2,\kf_3;\kf_4,\kf_5,\kf_6) =
  \fr{g^4}{8} \times}\nn \\ 
&&[ \, G(123,-,456) \nn \\
&&-\,G(12,3,456) - G(13,2,456) - G(23,1,456) \nn \\
&&-\,G(123,4,56) - G(123,5,46) - G(123,6,45) \nn \\
&&+\,G(1,23,456) + G(2,13,456) + G(3,12,456) \nn \\
&&+\,G(123,45,6) + G(123,46,5) + G(123,56,4) \nn \\
&&+\,G(12,34,56) + G(13,24,56) + G(23,14,56) \nn \\
&&+\,G(12,35,46) + G(13,25,46) + G(23,15,46) \nn \\
&&+\,G(12,36,45) + G(13,26,45) + G(23,16,45) \nn \\
&&-\, G(1,234,56) -  G(2,134,56) - G(3,124,56) \nn \\
&&-\, G(1,235,46) - G(2,135,46) - G(3,125,46) \nn \\
&&-\, G(1,236,45) - G(2,136,45) - G(3,126,45) \nn \\
&&-\, G(12,345,6) - G(12,346,5) - G(12,356,4) \nn \\ 
&&-\, G(13,245,6) - G(13,246,5) - G(13,256,4) \nn \\
&&-\, G(23,145,6) - G(23,146,5) - G(23,156,4) \nn \\
&&+\, G(1,2345,6) + G(2,1345,6) + G(3,1245,6) \nn \\
&&+\, G(1,2346,5) + G(2,1346,5) + G(3,1246,5) \nn \\
&&+\, G(1,2356,4) + G(2,1356,4) + G(3,1256,4) ] 
\,.
\eea
It is straightforward to prove this using the definition 
of $G$, see eq.\ (\ref{gdef}). After some cancellations 
one obtains the original form of the vertex in terms 
of the integrals $a,b,c,s$ and $t$ as it was given in \cite{BE}. 

With the new representation for $W\phi_2$ and thus 
for the two--to--six gluon vertex $V_{2 \rightarrow 6}$ 
we have established 
the conformal invariance of the latter in impact parameter 
space. With this result we can now conclude that 
the whole set of amplitudes of the GLLA with up to 
six $t$-channel gluons is conformally invariant. 
This is possible because in \cite{BE} it was shown 
that these amplitudes consist of only the following 
building blocks: states of $n$ reggeized gluons with 
$n=2,4,6$, two--to--four transition vertices that 
can be expressed in terms of $V$, and a two--to--six 
transition given in terms of $W$. All of these elements, 
including $W$, are conformally invariant in impact 
parameter space. 

A closer look at the expressions (\ref{vmitg}) and 
(\ref{wmitg}) reveals 
an interesting regularity in the construction of the 
functions $V$ and $W$ which define the two--to--four 
and the two--to--six gluon vertex, respectively. 
In the remaining part of this section we want to study 
this regularity. 

For simplicity of notation we again consider the action 
of $V$ and $W$ 
on the two--gluon state $\phi_2$. We recall that 
the function $(V\phi_2)(\kf_1,\kf_2;\kf_3,\kf_4)$ 
is symmetric in its first two arguments as well as 
in its last two arguments. This matches the symmetry of 
the color structure it comes with in the two--to--four 
vertex $V_{2 \rightarrow 4}$, see eq.\ (\ref{colV}). 
Similarly, the function $(W\phi_2)(\kf_1,\kf_2,\kf_3;\kf_4,\kf_5,\kf_6)$ 
is completely symmetric in its first three as well as in its last 
three arguments. Again, this matches the symmetry of the color structure with 
which this permutation of arguments occurs in the full vertex, 
see eq.\ (\ref{newpiece}). 

Let us now look in more detail at the arguments of the $G$-functions 
as they appear in the new representation for $W\phi_2$ 
in eq.\ (\ref{wmitg}). The first term is $G(123,-,456)$ which equals 
$({\cal K} \phi_2) (123,456)$ (up to a factor $N_c$). 
All other terms in eq.\ (\ref{wmitg}) can now 
be constructed from this term by a simple procedure. 
For each possible subset of the six momenta $\kf_i$ we obtain exactly 
one term. These respective terms have the sum of the momenta 
in that subset as the second argument of the function $G$. At the same 
time, those momenta are taken out of the sums in the 
first and third argument of $G$. The sign of the resulting term 
alternates with the number of elements, i.\,e.\ the sign 
equals $(-1)^l$ with $l$ being the number of elements of the respective 
subset of momenta. 

It appears useful to go through the above procedure in some examples. 
The simplest non--trivial subset of the six momenta is one containing 
only one element, say $\kf_1$. Starting from $G(123,-,456)$ we put 
$\kf_1$ into the second argument of $G$ while removing it from the 
first argument. With the minus sign ($l=1$, see above) we have 
$- G(23,1,456)$. Similar terms are obtained for the other five momenta. 
There are $15$ subsets of the six momenta that contain exactly two momenta. 
Let us consider the subset $\{ \kf_1,\kf_4\}$. Removing $\kf_1$ from the 
first and $\kf_4$ from the third argument of $G(123,-,456)$ and 
taking their sum as the second argument we have $G(23,14,56)$ 
which comes with a positive sign since now $l=2$. 
For $l \ge 3$ there are subsets which contains all of the momenta 
appearing in the first or third argument of $G(123,-,456)$, 
for example $\{ \kf_1,\kf_2,\kf_3\}$. In these cases the resulting 
$G$-function would have a vanishing first (or third) argument, and thus 
vanish itself due to (\ref{gwird0}). 

It is easily checked that this procedure reproduces the correct sign 
and momentum structure of all terms in the new 
representation (\ref{wmitg}) of the function $W \phi_2$. 
Let us recall that the term that we started with, $G(123,-,456)$, 
has exactly the same symmetry properties as the full 
function $W \phi_2$. 
This is consistent with the fact that the above procedure is 
completely symmetric in the six gluon momenta.

The function $V \phi_2$ as given in eq.\ (\ref{vmitg}) 
can be obtained by the same procedure, now starting from 
$G(12,-,34)$. We have thus found a general rule 
for constructing the known two--to--four and two--to--six 
gluon vertices in the effective field theory of unitarity corrections. 
The starting points are the terms $G(12,-,34)$ and $G(123,-,456)$, 
respectively. Moreover, that rule even works for the special 
case of the two--to--two transition vertex as which the BFKL kernel 
can be interpreted. In that case the term to start with is $G(1,-,2)$ 
and one can immediately apply eq.\ (\ref{gwirdbfkl}). 

Remarkably, this procedure leads to vertex functions 
$V \phi_2$ and $W \phi_2$ which vanish if {\em any} outgoing momentum 
vanishes. For $V \phi_2$ we have seen this already 
in eq.\ (\ref{nullstellv}). Similarly, for $W \phi_2$ we have 
\be
\left.
(W \phi_2) (\kf_1,\kf_2, \kf_3;\kf_4,\kf_5,\kf_6) 
\right|_{\kf_i=0} = 0 
\;\;\;\;\;\;  (i\in\{1,\dots,6\}) 
\,.
\label{nullstellw}
\ee
This is in contrast to the function $G$ from which they are 
constructed, see eq.\ (\ref{gwirdbfkl}). 
In \cite{Reggeization} it was discussed that this condition 
appears to be characteristic for all possible transition vertices of the 
effective field theory of unitarity corrections. 

It is now very suggestive to guess that a potential 
two--to--eight gluon vertex $V_{2 \rightarrow 8}$ 
would be constructed in a similar way. Obviously, the corresponding 
color structure --- the analogue of (\ref{newpiece}) --- 
is unknown. The corresponding color 
decomposition will presumably lead to a function $X$ with 
symmetry properties similar to $V$ and $W$, namely 
$X$ will be symmetric in its first four momentum 
arguments and in its last four momentum arguments. 
Its conjectured decomposition in terms of $G$-functions 
is obtained by the above rule starting from $G(1234,-,5678)$. 
We expect that the effective field theory of unitarity corrections 
contains further vertices $V_{2 \rightarrow 2m}$ ($m \in \N$) to 
which the same procedure applies. 

The computation of such vertices using the conventional 
method (see \cite{BE}) is extremely tedious. It would 
therefore be very interesting if our conjecture could be 
tested using other approaches to the problem of perturbative 
unitarity corrections. 

We should add that the rule given here fixes the vertex functions 
only up to their overall normalization. 
The overall factors of the coupling constant $g$ 
are adjusted in such a way that the power of $g$ equals 
the number of outgoing gluons (having in mind that $G$ already 
contains a factor $g^2$). The numerical factors of $1/2$ and 
$1/8$ in eqs.\ (\ref{vmitg}) and (\ref{wmitg}), however, cannot be 
explained unambiguously at the moment. 
Another very important piece of information that can at the moment 
only be obtained by the full computation of the vertices is their 
color structure. 

\section{Summary}

The perturbative unitarity corrections in the GLLA can be 
cast into the form of an effective field theory of interacting $n$-reggeized 
gluon states in the $t$-channel and transition vertices connecting 
states with different numbers of reggeized gluons. We have studied 
one of the building blocks of this effective field theory, namely 
the two--to--six reggeized gluon vertex. A new representation 
for this vertex was found which proves its invariance under 
global conformal (M\"obius) transformations in two--dimensional 
impact parameter space. With the help of this representation 
we have also been able to find a regularity in the expressions 
for the two--to--four and the two--to--six transition vertices. 
This regularity might be a helpful observation for the further investigation 
of the elements of the effective field theory of unitarity corrections. 

By proving the conformal invariance of the two--to--six transition 
vertex we have established that the full set of 
amplitudes in the GLLA with up to 
six reggeized gluons in the $t$-channel is conformally invariant. 
This lends further support to the conjecture that the whole set 
of unitarity corrections in the GLLA can be formulated as 
a conformal field theory in two--dimensional impact parameter 
space with rapidity as an additional real parameter. 
The still open and very challenging problem is of course 
to identify this conformal field theory. 

Eventually one would like to consider the unitarity corrections 
in next--to--leading logarithmic approximation (NLLA). 
So far only the NLL corrections to the BFKL Pomeron are known 
\cite{FL,CC}. One expects that the unitarity 
corrections will exhibit the structure of an effective field 
theory also in the generalized NLLA. It would be desirable 
to compute the NLL corrections to all elements of 
the effective field theory. This is an extremely difficult 
problem, and so far the only step in this direction has been 
to make an educated guess for the two--to--four gluon 
vertex in NLLA \cite{CE}. 
It is known that the NLL corrections to the BFKL Pomeron 
break the conformal invariance in impact parameter space. 
But it turns out that this breaking is soft in the sense that 
it occurs only due to the running of the gauge coupling. 
One hopes that this will be true also for the other elements 
of the unitarity corrections in the generalized NLLA.  
In this case the conformal invariance would remain an 
extremely powerful tool for the investigation of 
unitarity corrections in high energy QCD. 

\section*{Acknowledgements}
I would like to thank Jochen Bartels and Gian Paolo Vacca 
for helpful discussions.

\end{document}